\documentclass[fleqn,usenatbib]{mnras}

\usepackage{newtxtext,newtxmath}

\usepackage[T1]{fontenc}

\DeclareRobustCommand{\VAN}[3]{#2}
\let\VANthebibliography\thebibliography
\def\thebibliography{\DeclareRobustCommand{\VAN}[3]{##3}\VANthebibliography}

\usepackage{graphicx}	
\usepackage{amsmath}	
\usepackage{multirow}
\usepackage{tablefootnote}
\usepackage{threeparttable}
\usepackage{graphicx}
\usepackage{color}
\usepackage{booktabs}

\newcommand{\hii}{H\,{\sc ii} }

\newcommand{\oiii}{[O\,{\sc iii}] }
\newcommand{\oii}{[O\,{\sc ii}] }

\title[JWST NIRSpec/G395H IFU study of Gz9p3]{Diverse ionized gas conditions in a dynamically hot, interacting galaxy at $z=9.31$ revealed by JWST/NIRSpec IFU}

\author[Ikeda et al.]{
Ryota Ikeda,$^{1,2}$\thanks{E-mail: ryota195ikeda@gmail.com}
Takahiro Morishita,$^{2}$ Massimo Stiavelli,$^{3,4,5}$,
and Antonello Calabr\`{o}\,$^{6}$ \\
$^{1}$\,Leiden Observatory, Leiden University, P.O. Box 9513, 2300 RA Leiden, The Netherlands\\
$^{2}$\,Astronomical Institute, Tohoku University, 6-3 Aramaki, Aoba-ku,
Sendai 980-8578, Japan\\
$^{3}$\,Space Telescope Science Institute, 3700 San Martin Drive, Baltimore, MD 21218, USA\\
$^{4}$\,The William H. Miller III, Dept. of Physics \& Astronomy, Johns Hopkins University, Baltimore, MD 21218, USA\\
$^{5}$\,Dept. of Astronomy, University of Maryland, College Park, MD 20742, USA\\
$^{6}$\,INAF--Osservatorio Astronomico di Roma, via Frascati 33, 00078, Monte Porzio Catone, Italy
}

\date{Accepted XXX. Received YYY; in original form ZZZ}

\pubyear{\the\year{}}
\begin{document}
\label{firstpage}
\pagerange{\pageref{firstpage}--\pageref{lastpage}}
\maketitle

\begin{abstract}
We present a spatially and spectrally resolved study of an interacting galaxy, Gz9p3 at $z=9.31$, using the James Webb Space Telescope NIRSpec/G395H IFU observation with high-spectral resolution ($R\approx 2700$) mode. Gz9p3 consists of two sub-regions (`core' and `tail'), which show stark contrast in their physical properties. The \hii regions in the core are characterized by high electron density ($n_{e}\gtrsim2900\,{\rm cm}^{-3}$), low gas-phase metallicity ($0.25$\,dex below the mass-metallicity relation), and yet relatively low specific star formation rate (sSFR) among the system. On the other hand, the tail exhibits low electron density ($n_{e}<70\,{\rm cm}^{-3}$), relatively high gas-phase metallicity that is consistent with the mass-metallicity relation, and high sSFR. The integrated spectrum thus shows the properties inbetween these two regions. No evidence of ordered rotation in ionized gas is found, suggesting that Gz9p3 is a dynamically hot system. Finally, we find ionized gas outflows characterized by the secondary [O\,{\sc iii}]\,5007\AA\,\,line component throughout most of the system. The outflow velocity is below the escape velocity, making Gz9p3 one of the first galactic systems experiencing a galactic fountain.
Overall, our findings indicate that this system is in a phase in which the pristine gas is falling efficiently into the core, diluting the gas metallicity, and enhancing star formation and outflows after the galaxy interaction.
\end{abstract}

\begin{keywords}
galaxies: high-redshift -- galaxies: star formation -- galaxies: evolution -- galaxies: ISM -- ISM: H\,{\sc ii} regions
\end{keywords}



\section{Introduction}
\label{sec:section1}

Galaxy formation in the first $\sim500$\,Myr of the Universe has been found to proceed remarkably rapidly \citep{2025NatAs...9.1134A}. Detections of metal emission lines in multiple ultraviolet (UV)-bright galaxies at $z\gtrsim10$ (e.g., \citealp{2023Natur.622..707A}; \citealp{2023ApJ...957L..18S}; \citealp{2024A&A...690A.288B}; \citealp{2024ApJ...972..143C}; \citealp{2025NatAs...9..155Z}; \citealp{2025A&A...696A..87C}; \citealp{2025ApJ...988...19S};  \citealp{2026OJAp....956033N}; \citealp{2026OJAp....955261W}; \citealp{2026arXiv260708749L}), thanks to the high sensitivity of the Atacama Large Millimeter/submillimeter Array (ALMA) and the James Webb Space Telescope (JWST), indicate that the interstellar medium (ISM) of these galaxies has already been polluted by the first-generation massive stars. Such earliest galaxies are compact ($\lesssim500$\,pc in radius), and taken together with the velocity dispersion and stellar mass, their properties can be considered as a scaled-up version of ultra-compact dwarfs or nucleated dwarf galaxies in the nearby Universe (\citealp{2024ApJ...977L...9Z}; \citealp{2026OJAp....955261W}). 

Despite of active star formation, no direct detection of far-infrared (FIR) continuum emission in such UV-bright galaxies have been reported thus far (e.g., \citealp{2026ApJ..1000..159M}; \citealp{2026MNRAS.546f2284B}), leaving MACS0416-Y1 at $z=8.31$ as the most distant galaxy with FIR continuum emission originating from dusty ISM \citep{2019ApJ...874...27T}. Below $z\sim8.5$, many massive galaxies start to exhibit clumps or extended stellar structure, as well as dust continuum emission, which is in stark contrast to compact and UV-bright galaxies at $z\gtrsim10$. Simultaneously, this is the epoch when both theoretical simulations and observations report the emergence of dynamically cold gas disks (\citealp{2024A&A...685A..72K}; \citealp{2024MNRAS.535.2068R}).

In this context, the redshift range of $z\sim8.5-10$ (corresponding to a duration of 100\,Myr in cosmic age) is considered a transitional epoch when most massive galaxies have experienced the first peak of bursty star formation activity and begun to grow dust grains (\citealp{2025A&A...699A.336B}; \citealp{2026OJAp....959986N}). It is therefore important to investigate how the ISM conditions, particularly the ones of \hii regions, differ from the ones in galaxies at later epochs. 

Rest-frame UV-to-optical spectroscopic information is only available for such young, FIR-dark galaxies, and properties like electron density, gas-phase metallicity, and ionization parameter have been critical probe for understanding their ISM conditions. Well-established relations -- not only for integrated systems but also on spatially resolved scales -- are known for nearby to cosmic noon galaxies, namely mass-metallicity (e.g., \citealp{2020MNRAS.491..944C}; \citealp{2021ApJ...914...19S}; \citealp{2022MNRAS.510..320F}) and star formation rate (SFR) density-electron density (e.g., \citealp{2021ApJ...909...78D}; \citealp{2025ApJ...993L..51L}) relations, which reveal the evolution of the interplay between star formation activity, gas density, and gas-phase metallicity across cosmic time. Whether these correlations are already in place in such young galaxies -- if not, when they become established -- is of great interest.

Numerous studies have reported a statistical investigation on high-redshift galaxies (e.g., \citealp{2023ApJS..269...33N}; \citealp{2024A&A...684A..75C}; \citealp{2024ApJ...971...43M}; \citealp{2026ApJ..1001...38T}), revealing a large source-to-source variation in the early Universe. However, these studies mostly exploit the slit spectroscopy, which only covers the fraction of galaxies. Therefore, it remains largely unexplored whether the correlations hold true and how much they vary within a single object, which can only be reliably explored either by slit-stepping (e.g., \citealp{2025ApJ...983..139B}) or integrated field unit (IFU) observations.

\subsection{Gz9p3: an interacting galaxy with substructure at $z=9.3$ }
\label{subsec:section1.1}

In this paper, we conduct a spatially- and spectrally-resolved study of Gz9p3/DHZ1
(hereafter Gz9p3 following \citealp{2024NatAs...8..657B}), using the NIRSpec Integral Field Unit (IFU) high-spectral resolution mode ($R\approx2700$; \citealp{2022A&A...661A..80J}; \citealp{2022A&A...661A..82B}) on board the JWST. The coordinate of this system is (RA, Dec) $= (3.61716,\, -30.42555$) in degrees. This galaxy represents one of the most massive ($\log({M_{\star}/M_{\odot}})=9.20^{+0.14}_{-0.11}$) and extended systems known in the early Universe, spanning $\sim2$\,kpc in the source plane, making it an ideal target for IFU studies. At rest-frame UV wavelengths imaged by NIRCam F150W and F200W filters, the central core of Gz9p3 exhibits two distinct clumps, suggesting an ongoing merger \citep{2024NatAs...8..657B}. 

Gz9p3 was originally identified as a photometric high-redshift galaxy candidate with the Hubble Space Telescope (\citealp{2016A&A...590A..31C}; \citealp{2023ApJ...948L..14C}). The galaxy was then spectroscopically confirmed by \cite{2024NatAs...8..657B} as $z_{\rm spec}=9.3127\pm0.0002$ from NIRSpec Mirco Shutter Array (MSA) observations, followed by \cite{2024ApJ...977..250F} with the NIRSpec PRISM spectroscopy. This spectroscopic redshift marks at the edge of high-redshift limit where the [O\,{\sc iii}]\,5007\AA\, line emission falls into the NIRSpec G395M and G395H gratings. 

ALMA Band\,7 observations toward this object was reported by \cite{2026OJAp....965333A}. The [O\,{\sc iii}]\,88$\mu$m line emission is detected at $\sim6\sigma$ significance, making Gz9p3 one of the most distant sources with a FIR line detection. On the other hand, no dust continuum emission at $\lambda_{\rm rest}=88$\,$\mu$m is detected in Gz9p3, providing an upper limit on dust mass of $\log(M_{\rm dust}/M_{\odot})<6.28$, assuming a dust temperature of $T_{\rm dust}=50$\,K. \cite{2026arXiv260422460B} conduct NIRSPec IFU (G395M grating) and MIRI MRS observations, reporting both [O\,{\sc iii}]\,5007\AA\, and H$\alpha$ line emission. No evidence of active galactic nuclei (AGN) is found thus far.

Recently, Gz9p3 was reported to have super-solar carbon abundance though the inspection of the UV absorption line features (\citealp{2026arXiv260613078P}; \citealp{2026arXiv260608782N}; \citealp{2026arXiv260421516C}; \citealp{2026arXiv260421218Z}; \citealp{2026arXiv260809813W}).
For the observed signature of outflows induced by stellar winds and weak H$\beta$ line emission, \cite{2026arXiv260421516C} argue that Gz9p3 has likely experienced bursty star formation over the past $\sim10-20$\,Myr, followed by a declining star formation history (SFH).

Throughout this work, we adopt the magnification factor of $\mu_{\rm lens}=1.66\pm0.02$ reported by \cite{2023ApJ...948L..14C}. We further adopt a flat $\Lambda$CDM cosmology and the cosmological parameters of $H_{0}=70$\,km\,s$^{-1}$\,Mpc$^{-1}$, $\Omega_{\rm M}=0.3$, and $\Omega_{\rm \Lambda}=0.7$. At the redshift of $z = 9.313$ with $\mu_{\rm lens}=1.66$, a projected physical scale of 1\,arcsec corresponds to $4.365/\sqrt{\mu_{\rm lens}}\approx3.39$\,kpc. The cosmic age corresponds to $\approx510$\,Myr. All of the fluxes and spatial extent presented in this paper are corrected for lensing magnification. For the calculation of intrinsic fluxes, uncertainties in the lensing magnification are propagated into the flux uncertainties.
We adopt the Chabrier initial mass function (IMF; \citealp{2003PASP..115..763C}) when referring to or deriving stellar masses and SFRs. We consistently use vacuum wavelengths for emission lines throughout our analyses.

\section{Observations and Data Reduction}
\label{sec:section2}

\subsection{Observations}
\label{subsec:section2.1}

The NIRSpec IFU observations were performed on June 13, 2025 as part of the JWST Guaranteed Time Observations (GTO) program \#\,4553 (PI: M. Stiavelli). The high-resolution G395H/F290LP grating is used with the CYCLING dithering pattern. The exposure consists of 13 dithers, 22 groups per integration, and one integration per exposure. The NRSIRS2 readout pattern mode is used. The total exposure time is 5.85 hours (21,052 seconds). 

The galaxy is located in the UNCOVER field \citep{2024ApJ...974...92B}. We therefore retrieved the JWST/NIRCam images from the ASTRODEEP-JWST \citep{2024A&A...691A.240M}, which provides the HST and JWST imaging data that is reduced and combined self-consistently in this field across all available filters. We refer the reader to \cite{2024A&A...691A.240M} and the references therein for details of the observations and data reduction.

\subsection{Data Reduction}
\label{subsec:section2.2}

The data are reduced in the same way as in \cite{2025ApJ...985...83M}, using the official JWST pipeline (v1.20.2) with the Calibration Reference Data System (CRDS) v13.1.1 and a few additional custom steps. Briefly, we retrieve the level\,1 products (\_uncal.fits) from the Mikulski Archive for Space Telescopes (MAST). We run the stage\,1 step and then the stage\,2 step, with additional stripe elimination using {\tt NSClean} \citep{2025ApJ...978..108R} and cosmic-ray (CR) rejection using {\tt lacosmic} (\citealp{2001PASP..113.1420V}; \citealp{2023zndo..10145563B}). We also run the source detection tool {\tt SExtractor} \citep{1996A&AS..117..393B} in each \_rate image and flag significant point-like sources that are not detected in the previous steps as potential CRs. We run the stage\,3 step to construct a cube and subtract background in each wavelength frame using off-source regions. Lastly, we run {\tt lacosmic} again in each wavelength frame to mask any residual CRs.

Since we aim to conduct the spatially-resolved analyses, we perform an astrometric correction by comparing the NIRCam F356W image and NIRSpec data cubes. To this end, we constructed a continuum map by stacking all spectral channels within $\lambda_{\rm obs}=3.14-3.98$\,$\mu$m in the NIRSpec data cube, corresponding to the effective bandwidth of the F356W filter throughput curve. The choice of the F356W filter is made because no strong emission lines exist except for the \oii doublet and the throughput curve is entirely covered by the NIRSpec G395H/F290LP data cubes. The positional offset of ($\Delta{\rm R.A.}$, $\Delta{\rm Decl.}$) = ($+0\farcs129$, $+0\farcs147$) is found in the peak position of the continuum map with respect to that of the F356W image, and we applied an astrometric shift of ($-\Delta{\rm R.A.}$, $-\Delta{\rm Decl.}$) to the NIRSpec IFU data cube.

Finally, to compare the morphology and velocity structure of the rest-frame optical [O\,{\sc iii}] line with the FIR [O\,{\sc iii}] line reported in \cite{2026OJAp....965333A}, we reduced the ALMA Band\,7 data stored on the ALMA Science Archive. The data was taken as part of the Cycle\,11 program \#\,2024.1.00440.S (PI: H. Algera). We reduced the data by running the ALMA Pipeline (v6.6.1.17; \citealp{2023PASP..135g4501H}). We then create the synthesized cube images by adopting the CLEAN algorithm, using the {\tt tclean} task of the Common Astronomy Software Application package ({\tt CASA}; \citealp{2022PASP..134k4501C}). We adopt the Briggs weighting with the robust parameter $R=2.0$, and CLEANed down to $1.5$ times of the root mean square (rms) noise level of the dirty cube. The channel width is set to be 30\,km\,s$^{-1}$. The final synthesized beam size and the rms noise level are $0\farcs705\times0\farcs564$ (position angle $=84.2^{\circ}$) and 407$\,\mu$Jy\,beam$^{-1}$, respectively.

\begin{figure}
    \center
    \includegraphics[width=0.93\linewidth]{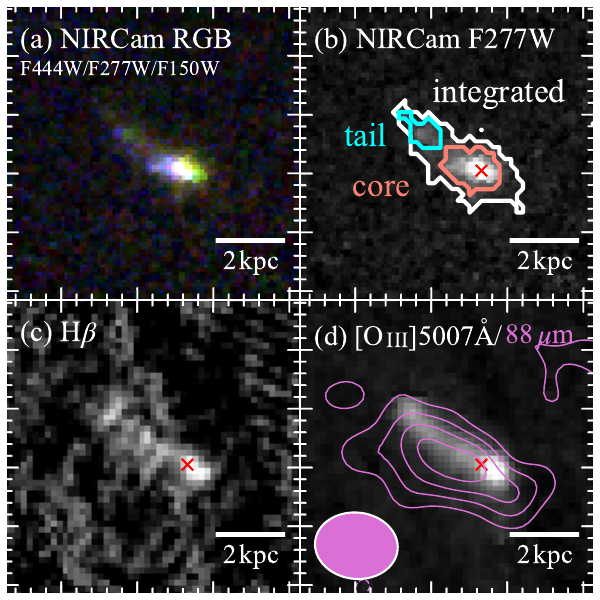}
    \caption{Cutout images ($2\farcs5$ on side) of Gz9p3. The scale bars at the lower right corner are corrected for lensing magnification. (a) The RGB (R: F444W, G: F277W, B: F150W) image. (b) NIRCam F277W image with three segmentation defined as `integrated' (white), `core' (red), and `tail' (cyan) regions. (c) Integrated H$\beta$ line emission. (d) Integrated [O\,{\sc iii}]5007\AA\, line emission. The purple contours display the integrated [O\,{\sc iii}]88$\mu$m line obtained by ALMA Band\,7 observations with contours starting from $2\sigma$ in steps of $1\sigma$. The red cross highlights the peak position in the F277W image.}
    \label{fig:figure1}
\end{figure}

\begin{figure*}
    \includegraphics[width=0.95\linewidth]{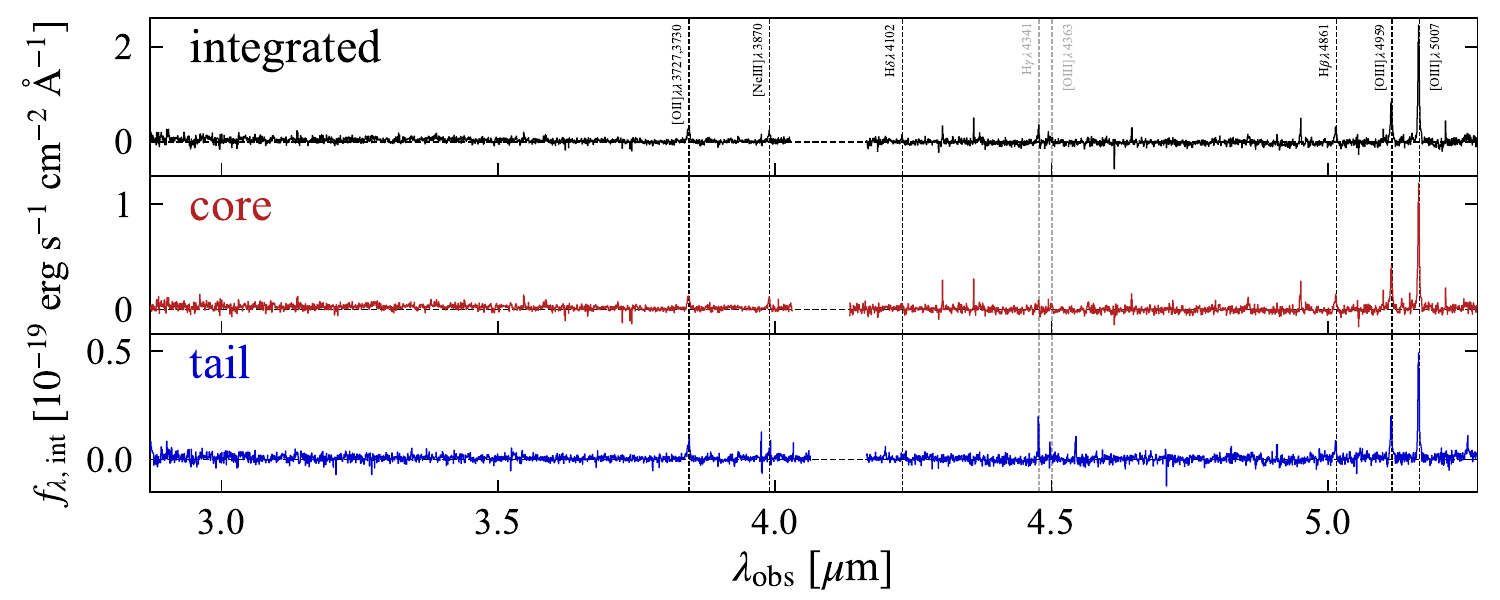}
    \caption{NIRSpec G395H/F290LP spectrum of integrated (top), core (middle), and tail (bottom) regions. The redshifted wavelengths of detected and non-detected ($<3\sigma$) lines are highlighted in black and grey dashed lines, respectively. The flux densities ($f_{\lambda, {\rm int}}$) are corrected for lensing magnification.}
    \label{fig:figure2}
\end{figure*}

\begin{figure*}
	\includegraphics[width=0.95\linewidth]{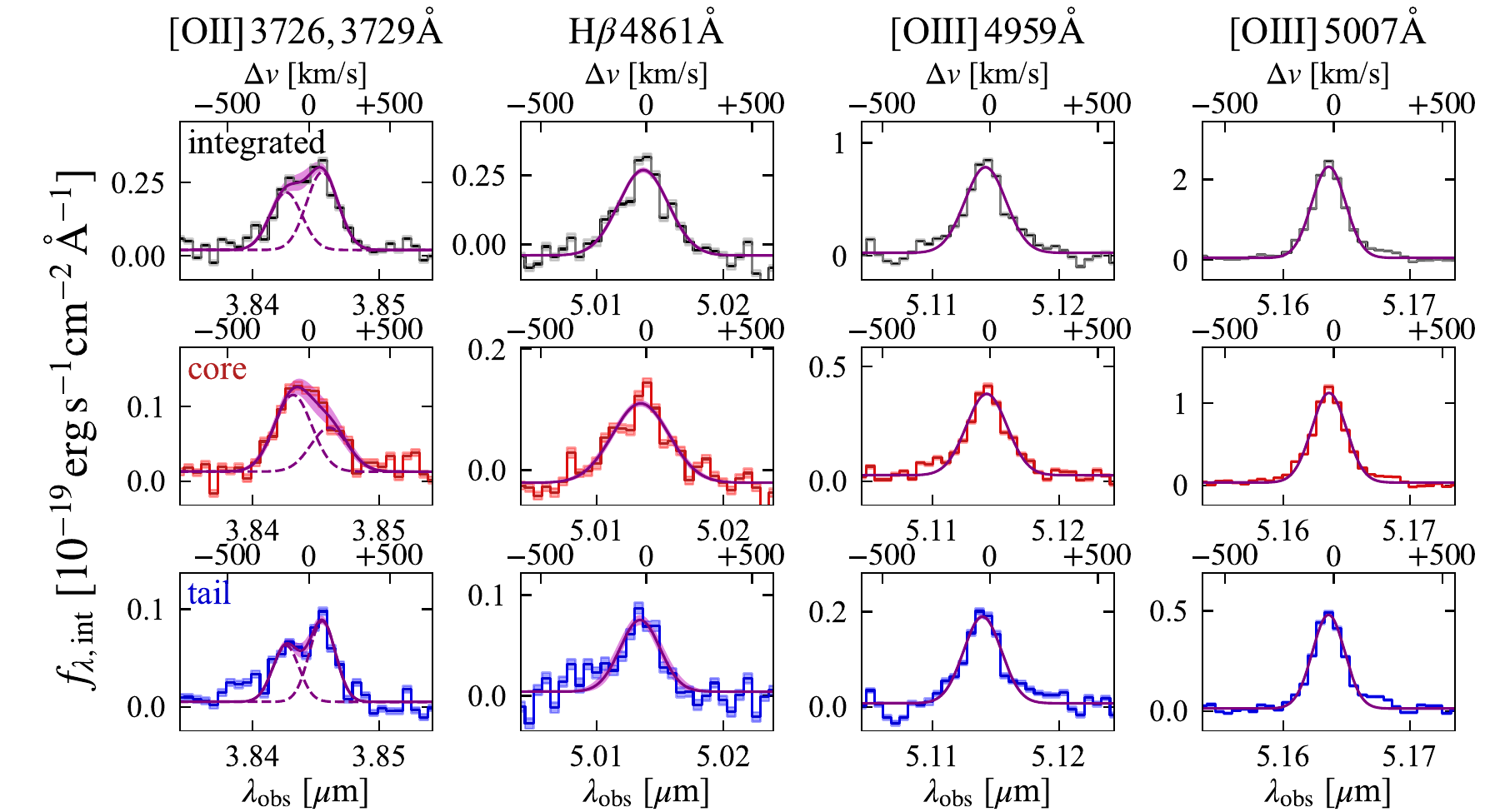}
    \caption{The zoom-in spectra of the \oii doublet, H$\beta$, [O\,{\sc iii}]4959\AA, and [O\,{\sc iii}]5007\AA\, lines. The shaded spectra show the flux uncertainties extracted from the ERR extension map. Each row presents the spectra extracted from the (sub)regions defined in Figure\,\ref{fig:figure1}. The best-fit Gaussian models are shown by the purple solid lines (median) and shaded areas (16th and 84th percentiles). The ticks on the top axis indicate the radial velocity offset relative to the wavelength of emission line at $z=9.311$.}
    \label{fig:figure3}
\end{figure*}

\section{Analyses and Results}
\label{sec:section3}

The major characteristics of Gz9p3 is the disturbed morphology with a tail-like sub-component extending toward the northeast from the brightest core. To investigate the spatial variation of physical properties in this system, we first run the {\tt SExtractor} \citep{1996A&AS..117..393B} using the stacked image of NIRCam long-wavelength filters (F277W, F356W, and F444W) and the NIRCam F277W image to define the integrated and sub-regions, respectively. Consequently, two sub-regions were identified: one associated with the brightest core (hereafter the `core' region) and the other associated with the tail-like feature (hereafter the `tail' region). Figure\,\ref{fig:figure1} and Figure\,\ref{fig:figure2} show the cutout images of Gz9p3 and the NIRSpec spectra extracted from different regions, respectively. The defined sub-regions are shown in the panel (b) of Figure\,\ref{fig:figure1}. The `integrated', `core', and `tail' regions have the surface area of 6.03\,kpc$^{2}$, 1.95\,kpc$^{2}$, and 0.98\,kpc$^{2}$, respectively.

The rest-frame optical nebular emission lines, namely the [O\,{\sc ii}]\,3726, 3729\AA, H$\beta$, [O\,{\sc iii}]\,4959, 5007\AA\, lines are detected in all regions. The intensity maps of H$\beta$ and [O\,{\sc iii}]\,5007\AA\, resemble the morphology of the NIRCam image, although we find a positional offset of $\sim0\farcs13=450$\,pc. We show the zoom-in spectra of the [O\,{\sc ii}]\,3726, 3729\AA, H$\beta$, and [O\,{\sc iii}]\,5007\AA\, lines with the best-fit Gaussian models in Figure\,\ref{fig:figure3}. The [O\,{\sc iii}]\,5007\AA\, spectra show a tentative excess in flux at the redshifted component ($\Delta v\sim250$\,km\,s$^{-1}$), which suggests the existence of outflow. We will discuss this further in Section\,\ref{subsec:section3.3.1}.

The [Ne\,{\sc iii}]\,3870\AA\, line is detected in both the `integrated' and `core' spectra, whereas the H$\delta$ line is detected only in the `integrated' spectrum. The H$\gamma$ and [O\,{\sc iii}]\,4363\AA\, lines are not detected at $>3\sigma$ significance in any of the spectra, either because the lines are intrinsically faint or because their spectral regions are affected by noisy pixels caused by cosmic rays. Given that the line flux of [O\,{\sc iii}]\,4363\AA\, line, which is crucial in determining the electron temperature though the `direct' method, is unavailable, we assume a constant electron temperature of $T_{e}=10^{4}$\,K, which is typical of H\,{\sc ii} regions, throughout this paper.

We describe our line fitting procedure and flux measurements in Appendix\,\ref{sec:sectionA}.

\subsection{Kinematical structure}

\begin{figure}
	\includegraphics[width=\linewidth]{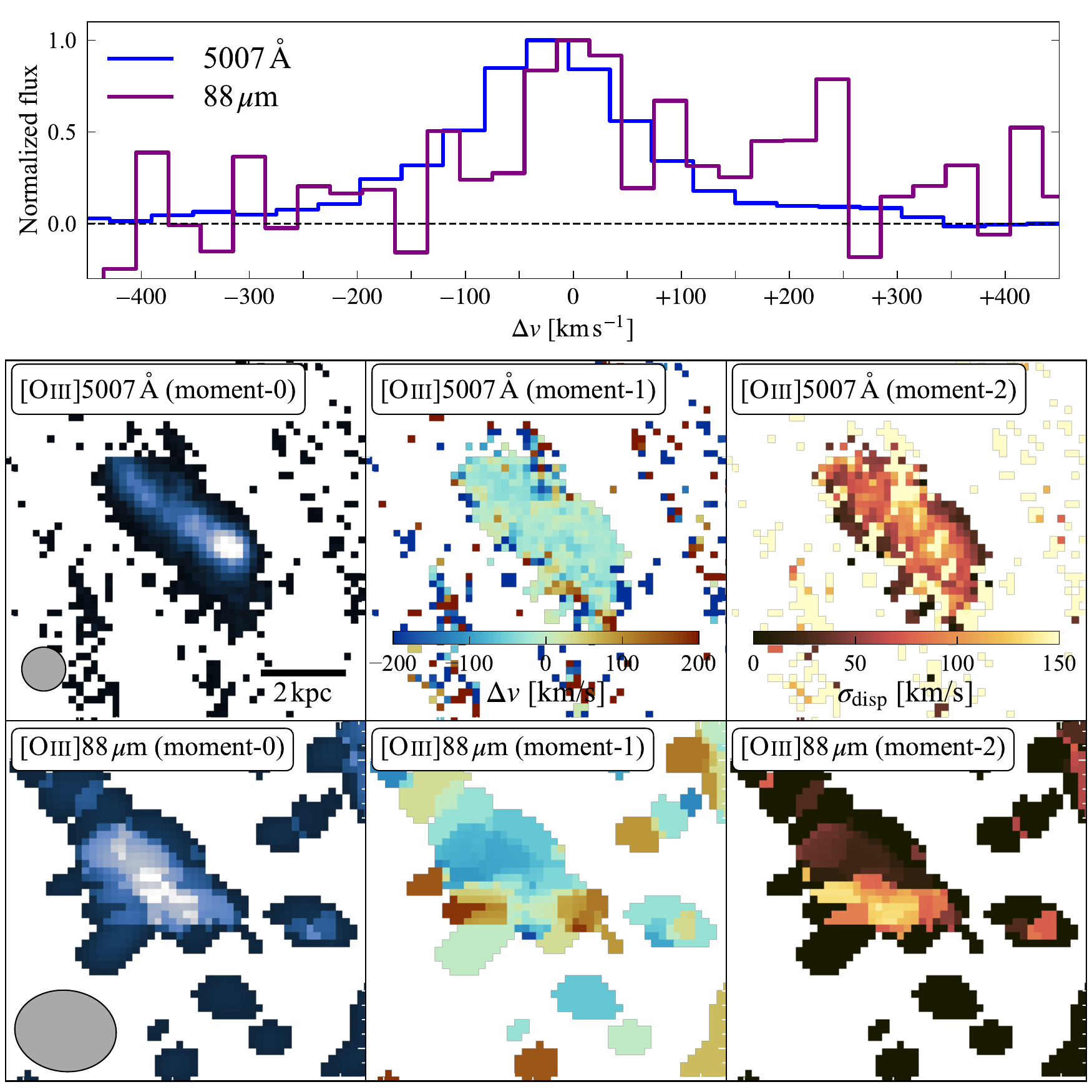}
	\caption{(Top) Comparison of the spectra between [O\,{\sc iii}]\,5007\AA\, (blue) and [O\,{\sc iii}]\,88$\mu$m (purple) lines. The reference velocity is aligned to the peak wavelength of the [O\,{\sc iii}]\,88$\mu$m line, which corresponds to $z=9.311$. (Bottom) The moment maps between the [O\,{\sc iii}]\,5007\AA\, and [O\,{\sc iii}]\,88$\mu$m lines. For the moment-1 maps, the reference velocity is the same as that used for the spectra shown at the top. For the moment-2 maps, the velocity dispersions are calculated relative to the peak wavelength of each emission line. Both the point spread function (PSF) of the JWST/NIRSpec IFU measured by \protect\cite{2024NatAs...8.1443D} and the ALMA synthesized beam are shown in the lower-left corner.}
    \label{fig:figure4}
\end{figure}

Figure\,\ref{fig:figure4} compares the spectra and moment maps between the [O\,{\sc iii}]\,5007\AA\, and [O\,{\sc iii}]\,88$\mu$m lines. To construct the velocity maps, we defined the reference velocity ($\Delta v = 0$) from the [O\,{\sc iii}]\,88$\mu$m line spectra extracted from $1''$ aperture. The peak velocity of the [O\,{\sc iii}]\,5007\AA\, line is consistent with the one of the [O\,{\sc iii}]\,88$\mu$m line.
We create the moment-0 (intensity), moment-1 (velocity), and moment-2 (velocity dispersion) maps of both \oiii lines defined as 
\begin{equation}
M_{n} = \cfrac{\sum_{i} f_{i}v_{i}^{n}}{\sum_{i} f_{i}},
\end{equation}
\noindent where $n$ is the moment order and $i$ is the pixel in the cube data. Here, we adopted pixel thresholds of $4\sigma$ and $2\sigma$ for constructing the moment maps of the [O\,{\sc iii}]\,5007\AA\, and [O\,{\sc iii}]\,88$\mu$m lines, respectively. 

As already indicated in Figure\,\ref{fig:figure1}, there is a spatial offset of $\sim0\farcs5$ between the [O\,{\sc iii}]\,5007\AA\, and [O\,{\sc iii}]\,88$\mu$m lines. While the [O\,{\sc iii}]\,5007\AA\, line emission peaks at the `core' region, the [O\,{\sc iii}]\,88$\mu$m line emission peaks at the midpoint of the `core' and `tail' regions. Although the angular resolution of the [O\,{\sc iii}]\,88$\mu$m map is more than two times coarser than the [O\,{\sc iii}]\,5007\AA\, map, the spatial offset is comparable to the ALMA beam size; thus, the spatial offset is likely to be real. 

The velocity (moment-1) maps show no clear evidence of ordered rotation in either the [O\,{\sc iii}]\,5007\AA\, or [O\,{\sc iii}]\,88$\mu$m line emission. The velocity dispersion (moment-2) of the [O\,{\sc iii}]\,88$\mu$m is highest ($\sigma_{\rm disp}\sim130$\,km\,s$^{-1}$) at the intensity peak, whereas the velocity dispersion of the [O\,{\sc iii}]\,5007\AA\, is relatively flat across the system. This non-rotating nature of Gz9p3 aligns with the picture that the system is in a phase of galaxy interaction, hampering to form a dynamically cold disk. This is also consistent with other galaxies at the similar epoch (\citealp{2023ApJ...952....9T}; \citealp{2024MNRAS.533.2488M}; \citealp{2026arXiv260514922T}; \citealp{2026arXiv260300232P}), suggesting that the ionized gas in massive star-forming galaxies at $z\sim8-10$ is generally dispersion-dominated ($\Delta v/\sigma_{\rm disp}\lesssim 2$) and dynamically hot. Future high-resolution follow-up of the [O\,{\sc iii}]\,88$\mu$m line emission can provide more precise kinematical structure at higher significance.

\begin{table}
	\centering
	\caption{Properties of both the integrated region and the subregions of Gz9p3.}
	\label{tab:table1}
    \renewcommand{\arraystretch}{1.2}
    \setlength{\tabcolsep}{2pt}
	\begin{tabular}{lccc} 
		\hline
		properties & integrated & core & tail \\
		\hline
        $M_{\star}$ [$10^{9}M_{\odot}$]\,$^{\rm a}$ & $1.6_{-0.4}^{+0.5}$ & $1.0_{-0.3}^{+0.3}$ & $0.17_{-0.04}^{+0.05}$ \\ 
        SFR\,(H$\beta$) [$M_{\odot}\,{\rm yr}^{-1}$] & $23.8\pm0.6$ & $11.5\pm0.4$ & $4.4\pm0.3$ \\
        sSFR [$10^{-8}{\rm yr}^{-1}$] & $1.48 \pm 0.37$ & $1.11 \pm 0.28$ & $2.58 \pm 0.64$ \\
        $\Sigma_{\rm SFR}$ [$M_{\odot}\,{\rm yr}^{-1}\,{\rm kpc}^{-2}$]$\,^{\rm b}$ & $3.9\pm0.1$ & $5.9\pm0.2$ & $4.5 \pm 0.3$ \\
        $n_{e}$\,([O\,{\sc ii}]) [cm$^{-3}$] & $63^{+54}_{-46}$ & $2870^{+1150}_{-1370}$ & $< 69$ \\
        $12+\log({\rm O/H})$$\,^{\rm c}$ & $7.54_{- 0.03}^{+0.03}$ & $7.52_{-0.04}^{+0.04}$ & $7.71_{-0.14}^{+0.11}$ \\ 
        $\log U$$\,^{\rm d}$ & $-2.43\pm0.01$ & $-2.36\pm0.02$ & $-2.52\pm0.01$ \\
        \hline
	\end{tabular}
    \begin{tablenotes}
    \item We use the {\tt ASYMMETRIC\_UNCERTAINTY} package \citep{2022ascl.soft08005G} to estimate the asymmetric uncertainties of the listed properties.
    \item $^{\rm a}$ The `integrated' stellar mass taken from \cite{2024NatAs...8..657B}. The stellar masses of the sub-regions are estimated from their fractions of the F444W flux, assuming a constant mass-to-light ratio.
    \item $^{\rm b}$ The SFR surface density defined as $\Sigma_{\rm SFR}={\rm SFR(H}\beta)/A$, where $A$ is the surface area of each region.
    \item $^{\rm c}$ R23-based gas-phase metallicity following the bet-fit relation presented by \cite{2025ApJ...985...24C}. 
    \item $\,^{\rm d}$ The ionization parameter $U$ derived from the photoionization model (with a metallicity of $Z=0.1Z_{\odot}$) presented by \cite{2019ApJ...874...93B}.
    \end{tablenotes}
\end{table}

\subsection{Physical Properties}
\label{subsec:section3.2}

In this section, we estimate and discuss the physical parameters, namely dust attenuation (Section\,\ref{subsec:section3.2.1}), gas-phase metallicity (Section\,\ref{subsec:section3.2.2}), electron density (Section\,\ref{subsec:section3.2.3}), and ionization parameter (Section\,\ref{subsec:section3.2.4}) for each sub-region. We summarize the measurements in Table\,\ref{tab:table1}. The extended nature of Gz9p3 allowed us to reveal the wide spatial variations across the system.

\subsubsection{Dust attenuation}
\label{subsec:section3.2.1}

Whether the ISM properties of Gz9p3 show signature of dust attenuation and thereby the dust built up is of considerable interest. From the spectral energy distribution (SED) fitting analysis, \cite{2024NatAs...8..657B} inferred the minimal stellar attenuation of $A_{V}\sim0.2$ with the UV slope $\beta=-1.94^{+0.05}_{-0.06}$. On the other hand, \cite{2026arXiv260608782N} reported a slightly higher value of $A_{V}= 0.48^{+0.09}_{-0.08}$. For nebular attenuation, \cite{2026arXiv260422460B} measured the Balmer decrement of H$\alpha$/H$\beta=2.4\pm0.5$ and H$\gamma$/H$\beta=0.3\pm0.1$ from the integrated IFU spectrum, which is consistent with no dust attenuation for the Case B recombination (\citealp{1989agna.book.....O}; see also \citealp{2026arXiv260421516C} for similar measurements). From our flux measurements, we obtained an updated decrement of H$\delta$/H$\beta=0.27\pm0.02$ and $0.26\pm0.03$ for the `integrated' and `core' regions, respectively, which is again consistent with the Case B recombination within the uncertainty. 

Taken together with the non-detection of dust continuum emission in the ALMA data ($\log(M_{\rm dust}/M_{\odot})<6.28$; \citealp{2026OJAp....965333A}), we conclude that nebular attenuation is negligible in this system. Therefore, we do not apply any dust attenuation correction to the nebular emission lines in the analyses hereafter.

\subsubsection{Gas-phase metallicity}
\label{subsec:section3.2.2}

We derived the gas-phase metallicity represented by the oxygen abundance. To account for different ionization phases of oxygen, we adopt the `strong line' method using the R23 line ratio defined as 
\begin{equation}
{\rm R23} = \cfrac{\lbrack {\rm O\,\textsc{ii}}\rbrack\lambda\lambda3727, 3729 + \lbrack {\rm O\,\textsc{iii}}\rbrack\lambda\lambda4959, 5007)}{{\rm H}\beta} ,
\end{equation}
which was recently calibrated from `direct' $T_{e}$-based metallicities by \cite{2025ApJ...985...24C}\,\footnote{We note that \cite{2025ApJ...985...24C} also reported a new empirical calibration $\hat{\rm R}$, defined as $\hat{\rm R}=0.18 \log({\rm R2}) + 0.98 \log({\rm R3})$, which reduces the scatter in the measurements of high-redshift galaxies around the best-fit correlation with the `direct' $T_{e}$-based metallicity. Following this calibration, we derived $12+\log({\rm O/H})= 7.56\pm0.02$, $7.55\pm0.03$, $7.68\pm0.08$ for the `integrated', `core', and `tail' regions, respectively, which are consistent with the $\rm R23$ measurements within the uncertainty.}. By using the best-fit correlation between the R23 line ratio and the gas-phase metallicity, we derived $12+\log({\rm{O/H}}) = 7.54_{- 0.03}^{+0.03}$, $7.52_{-0.04}^{+0.04}$, and $7.71_{-0.14}^{+0.11}$ for `integrated', `core', and `tail' regions, respectively. The metallicity of the `integrated' and `core' regions are in good agreement with the one derived by \cite{2024NatAs...8..657B}, who used the [Ne\,{\sc iii}]-to-[O\,{\sc ii}] line ratio (Ne3O2). The higher metallicity found in the `tail' region compared to the `core' region is at odds with the measurement from \cite{2026arXiv260422460B}, who found by $\sim0.5$\,dex lower metallicity in the tail component. While \cite{2026arXiv260422460B} derived gas-phase metallicity by adopting a non-parametric approach, using [O\,{\sc ii}], H$\beta$, [O\,{\sc iii}], and H$\alpha$ lines, we found a similar discrepancy in the R23 values between this work and \cite{2026arXiv260422460B}. Therefore, we suggest that this discrepancy is caused by the combination of differences in the spectral resolution between the PRISM ($R\approx100$) and high-resolution ($R\approx2700$) modes and by the methodology of deriving gas-phase metallicity. For consistency with the other literature (\citealp{2024ApJ...971...43M}; \citealp{2025ApJ...985...24C}), we adopt the measurements derived from the `strong line' method.

\begin{figure}
	\includegraphics[width=0.95\linewidth]{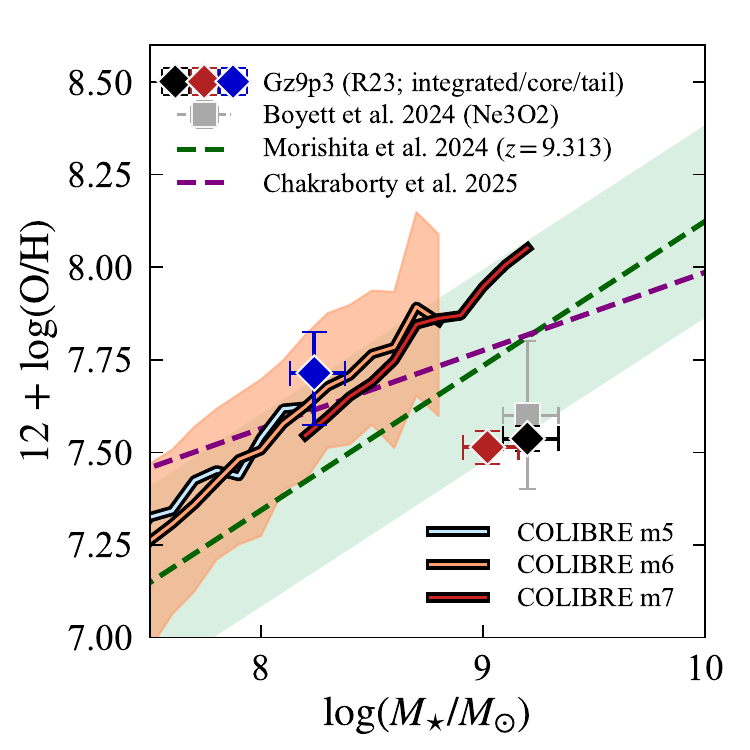}
	\caption{The stellar mass-metallicity relation (MZR) of Gz9p3. Green and purple dashed lines are the calibrated MZRs derived by \protect\cite{2024ApJ...971...43M} and \protect\cite{2025ApJ...985...24C}, respectively. The gray square shows the measurement from \protect\cite{2024NatAs...8..657B}. The three colored polylines are the MZR\ at $z=10$ reproduced by the COLIBRE simulation \protect\citep{2026arXiv260625995S}, with the uncertainty shown only for the intermediate resolution (m6).}
    \label{fig:figure5}
\end{figure}

Figure\,\ref{fig:figure5} shows the metallicity measurements in the context of the stellar mass-metallicity relation (MZR). Here, we estimated the stellar masses of the sub-regions by scaling the total stellar mass according to the F444W flux fraction, assuming a constant mass-to-light ratio. For comparison, three MZRs, two from observations (\citealp{2024ApJ...971...43M}; \citealp{2025ApJ...985...24C}) and one from the COLIBRE cosmological simulations \citep{2026arXiv260625995S} are shown. 

Compared to all of the MZRs in the literature, both `integrated' and `core' regions are located below the MZRs in the literature by $\sim0.25-0.5$\,dex which is consistent with \cite{2024NatAs...8..657B}, whereas the `tail' region is consistent or located slightly above the MZRs. We will discuss this result further in Section\,\ref{sec:section4}.

\subsubsection{Electron density}
\label{subsec:section3.2.3}

To estimate the electron densities, we measure the \oii line ratio obtained from the spectra. The line ratio of the two \oii lines comprising the doublet is known to primarily depend on the electron density and only weakly depend on the electron temperature. Since the electron temperature is not available for this object due to the non-detection of [O\,{\sc iii}]\,4363\,\AA, we opted to assume $T_{e}=10^{4}$\,K, which is the common temperature for \hii regions. We supplement the method of the double Gaussian fittings in Appendix\,\ref{subsec:sectionA1}. Following the formula presented in \cite{2016ApJ...816...23S}, we obtained $n_{e}= 63^{+54}_{-46}$\,cm$^{-3}$, $2870^{+1150}_{-1370}$\,cm$^{-3}$, and an upper limit of $<69$\,cm$^{-3}$ for `integrated', `core', and `tail' regions, respectively.

As a caveat, we note that recent studies have reported growing evidence that high-redshift galaxies at $z\gtrsim9$ observed thus far have typical electron temperatures of $T_{e}=(1.5-2.0)\times10^{4}$\,K (\citealp{2023ApJS..269...33N}; \citealp{2023ApJ...957L..18S}; \citealp{2024ApJ...973....8H}). By adopting the formula that takes into account the variations in electron temperature (\citealp{2019ARA&A..57..511K}; \citealp{2024ApJ...973...47A}), we found that increasing the electron temperature from $T_{e}=1.0\times10^{4}$\,K to $T_{e}=2.0\times10^{4}$\,K results in an increase in the electron density by $\sim0.14$\,dex (a factor of $\sim1.4$). Therefore, the choice of electron temperature has a minimal impact on our conclusions in this work. We also note that the uncertainties of the electron densities measured in this analysis are beyond the sensitivity of this variation.

\begin{figure}
	\includegraphics[width=0.95\linewidth]{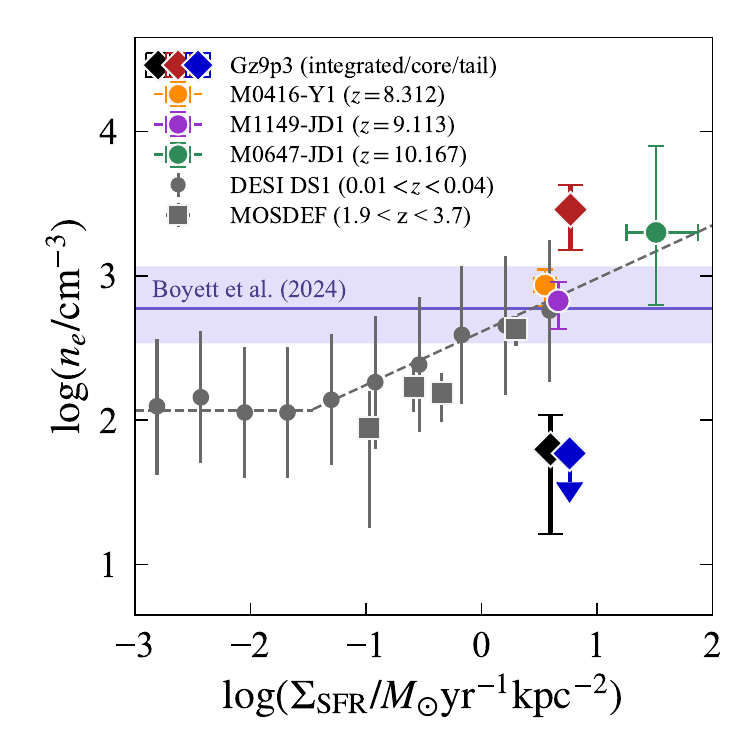}
	\caption{Electron density ($n_{e}$) as a function of SFR surface density ($\Sigma_{\rm SFR}$). The measurements for Gz9p3 is shown in diamond markers. The measurements from NIRSpec micro-shutter assembly (MSA) is highlighted by a purple horizontal band \protect\citep{2024NatAs...8..657B}. The local correlation derived from the DESI DR1 (\protect\citealp{2025ApJ...993L..51L}; after modifying for the \oii doublet from the [S\,{\sc ii}] doublet using the correlation by \protect\citealp{2026arXiv260628129R}), and high-redshift galaxies (the MOSDEF sample: \protect\citealp{2023ApJ...951...56R}, MACS0647-JD: \protect\citealp{2023ApJ...949L..34H}; \protect\citealp{2024ApJ...973...47A}, MACS1149-JD1: \protect\citealp{2024A&A...686A..85A}; \protect\citealp{2024MNRAS.533.2488M}, and MACS0416-Y1: \protect\citealp{2024ApJ...977L..36H};  \protect\citealp{2026arXiv260514922T}) are also shown for comparison.}
    \label{fig:figure6}
\end{figure}

In Figure\,\ref{fig:figure6}, we compare the SFR surface densities ($\Sigma_{\rm SFR}$) and electron densities ($n_{e}$) of the sub-regions in Gz9p3. To derive the SFR surface densities, we first calculated the H$\alpha$-based SFR \citep{1998ARA&A..36..189K} from the measured H$\beta$ line flux by assuming an intrinsic H$\alpha$/H$\beta$ ratio of 2.86. Then, we calculate the SFR surface densities by dividing the surface area comprising each sub-regions. 

\cite{2025ApJ...993L..51L} reported a correlation between [S\,{\sc ii}] doublet-based electron densities and SFR surface densities from a sample of 9590 star-forming galaxies at $0.01 < z < 0.04$. By applying the empirical conversion from [S\,{\sc ii}]-based electron density to [O\,{\sc ii}]-based electron density reported by \cite{2026arXiv260628129R}, we also show the $\Sigma_{\rm SFR}$-$n_{e}$ correlation in Figure\,\ref{fig:figure6}. Furthermore, we plot three high-redshift galaxies at $8\lesssim z\lesssim 10$ in Figure\,\ref{fig:figure6} for comparison, in which SFR surface density and electron density measurements are available (\citealp{2023ApJ...949L..34H}; \citealp{2024ApJ...973...47A}; \citealp{2024MNRAS.533.2488M}; \citealp{2024ApJ...977L..36H};  \citealp{2026arXiv260514922T}). If the SFR surface density is not explicitly reported in the literature, we adopted the effective radius ($R_{e}$) of H$\alpha$ line \citep{2024A&A...686A..85A} or rest-frame UV continuum \citep{2024ApJ...977L..36H} as a proxy of the spatial extent of star formation and derived the SFR surface density as $\Sigma_{\rm SFR}={\rm SFR}/2\pi R_{e}^{2}$. 

In the respect of the integrated measurements, Figure\,\ref{fig:figure6} demonstrates that most high-redshift galaxies follow the empirical trend of nearby galaxies between SFR surface density and electron density. However, we found that the integrated measurement of Gz9p3 has lower electron density whereas the `core' region has elevated electron density compared to the other high-redshift galaxies as well as the $\Sigma_{\rm SFR}-n_{e}$ correlation. The integrated measurement from \cite{2024NatAs...8..657B} is close to the one from the `core' region, which is reasonable, as the NIRSpec MSA slit was positioned over the `core' region. As we obtained only an upper limit for the `tail' region, it is likely that the `integrated' measurement represents an intermediate value between the `core' and `tail' regions. Overall, our electron density measurements provide a valuable insight at $z>9$, demonstrating that the electron density can vary by more than an order of magnitude within a single system.

\begin{figure*}
	\includegraphics[width=0.83\linewidth]{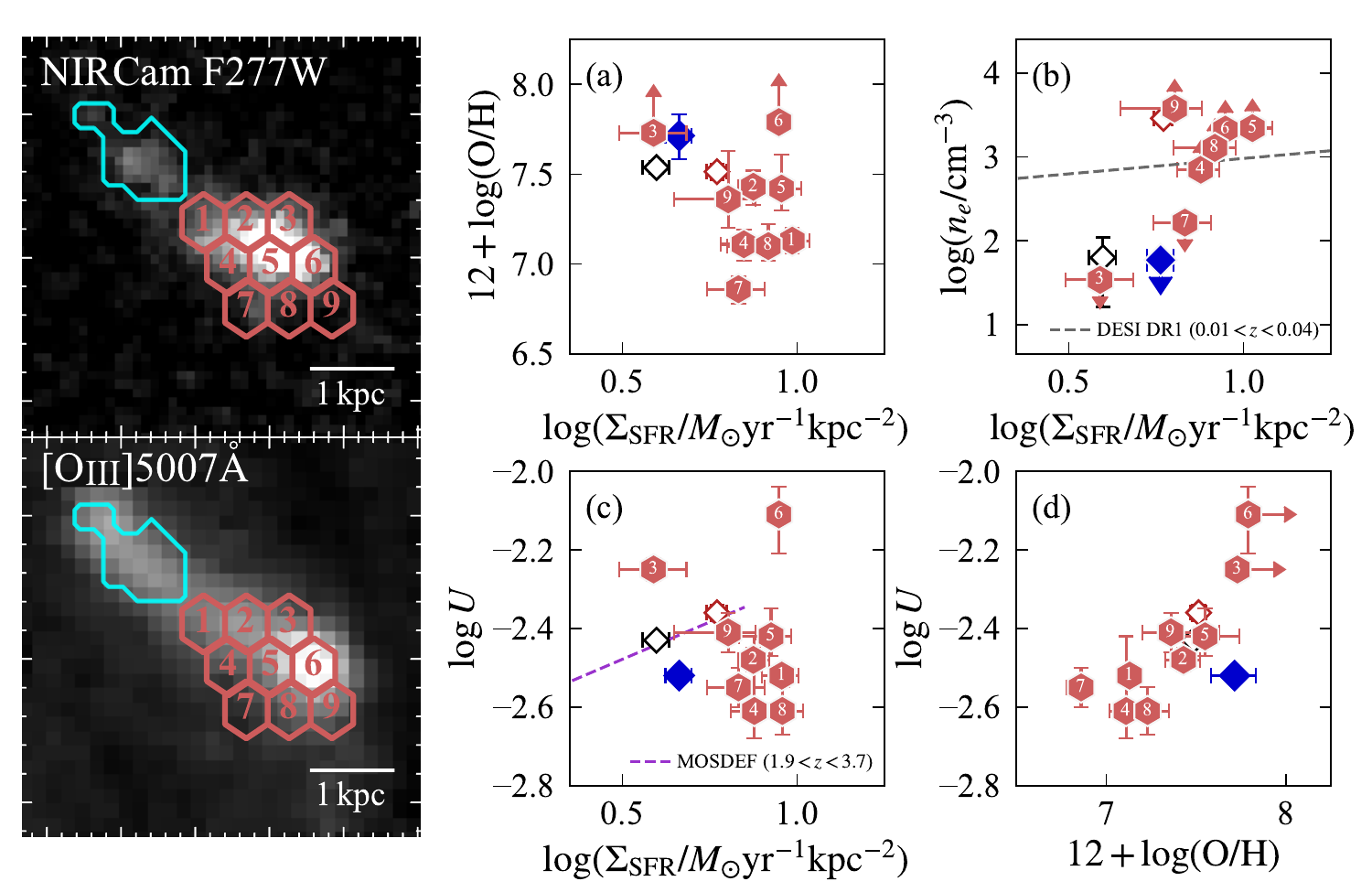}
	\caption{Spatially-resolved properties of the `core' region in Gz9p3. (Left) Positions of the nine hexagonal regions (labeled Regions 1 -- 9) overlaid on the NIRCam F277W image and the [O\,{\sc iii}]\,$5007$\AA\, line map. The `tail' region is also indicated by a cyan contour. (Right) Spatially-resolved correlations of (a) $\log\Sigma_{\rm SFR}$ - $12+\log({\rm O/H})$, (b) $\log\Sigma_{\rm SFR}$ - $\log n_{e}$, (c) $\log\Sigma_{\rm SFR}$-$\log U$, and (d) $12+\log({\rm O/H})$-$\log U$. The region number is labeled in each marker. The grey dashed line in panel (b) is the same as shown in Figure\,\ref{fig:figure6} from \protect\cite{2025ApJ...993L..51L}. The purple dashed line in panel (c) is the correlation reported by \protect\cite{2023ApJ...951...56R} as part of the MOSDEF Survey. The measurements for the `tail' region is shown as a blue diamond, while for the `integrated' and `core' regions are shown as open diamonds.}
    \label{fig:figure7}
\end{figure*}

\subsubsection{Ionization parameter}
\label{subsec:section3.2.4}

Finally, to derive the ionization parameter defined as $U=Q_{0}/(4\pi r^{2}n_{e}c)$, where $Q_{0}$ and $r$ are the number of ionizing photon emitted per unit time and the typical size of the \hii regions, respectively, we used the photoionization model presented by \cite{2019ApJ...874...93B}. Here, we used the correlation between the ionization parameter and the O32 ($=$ [O\,{\sc iii}]\,5007\AA/[O\,{\sc ii}]\,3726,\,3729\AA) line ratio. 

The derived ionization parameters are overall similar (with $0.2$\,dex variation) among the sub-regions. We note that the derived ionization parameters are consistent with the correlation between the [Ne\,{\sc iii}]/[O\,{\sc ii}] and O32 line ratios investigated by \cite{2021MNRAS.508.1686W}.

\begin{figure}
	\includegraphics[width=\linewidth]{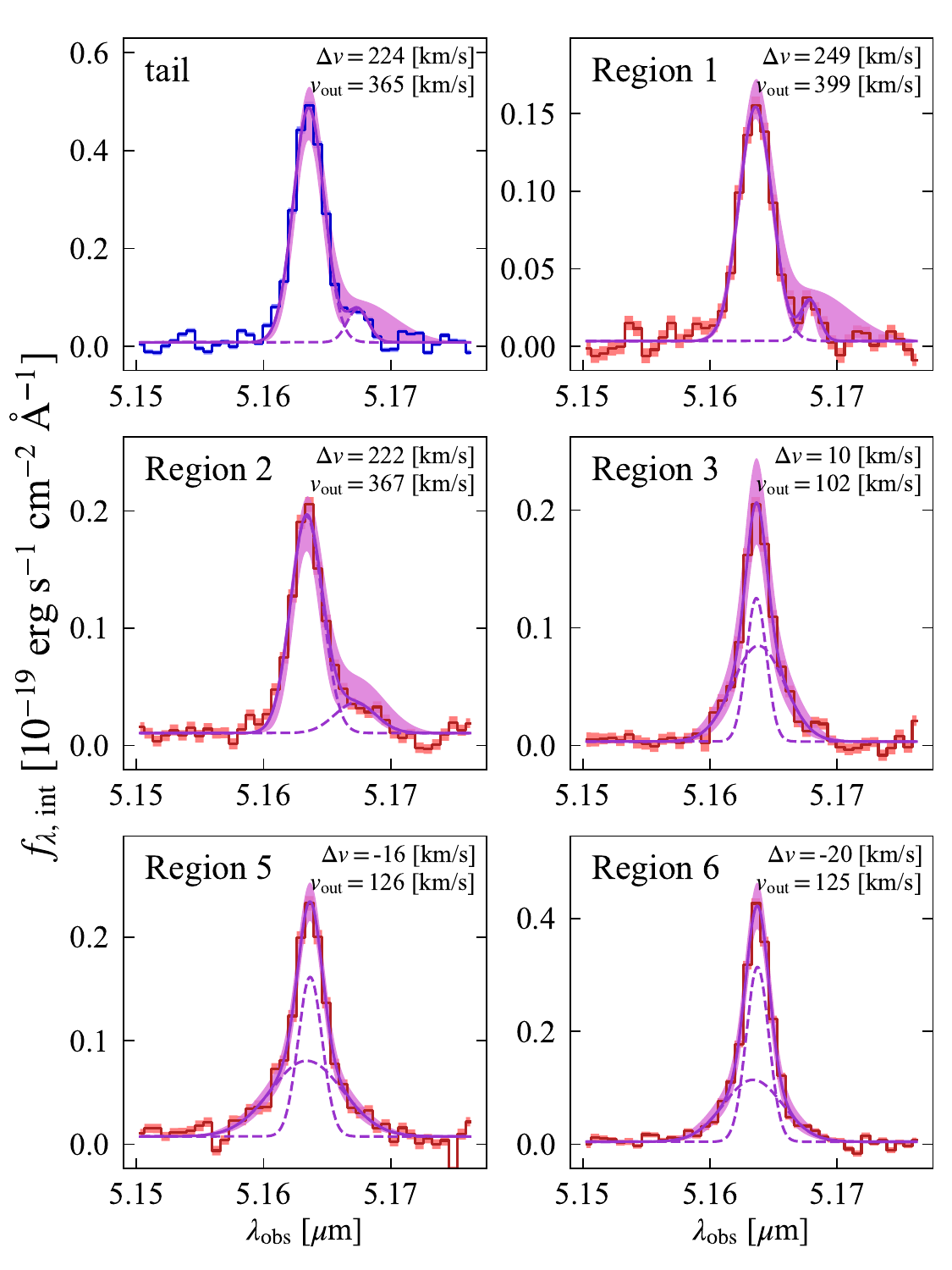}
    \caption{Examples of the [O\,{\sc iii}]\,5007\AA\, outflows in the `tail' and several hexagonal regions defined in Figure\,\ref{fig:figure7}. Outflows were characterized by performing double Gaussian fitting. Both velocity offset ($\Delta v$) between narrow and broad components and outflow velocity ($v_{\rm out}$) are shown in the top-right corner of each panel. Near the peak of the \oiii emission, the line profile exhibits an asymmetric broad wing. Toward the northeast, the redshifted component becomes more dominant.}
    \label{fig:figure8}
\end{figure}

\subsection{Spatially Resolved Analysis} 

In the previous section, we have investigated the physical properties of sub-regions, defined by the detection (segmentation) maps using {\tt SExtractor}. Although the extracted lines have high signal-to-noise ratio (S/N), the line ratios reflect the light-averaged value with possibly different ISM conditions over sub-kpc scales, making the interpretation uncertain. This has already been demonstrated by the wide variety of physical properties across the sub-regions derived in the previous section.

Therefore, to further assess the spatial variation of physical properties with finer spatial resolutions, we divided the `core' region into nine hexagonal regions (labeled Regions $1 - 9$) with a side length of $0\farcs1$ ($\sim340$\,pc). The diameter is larger than the PSF size of the NIRSpec IFU, thus each region is free from beam smearing caused by the others. The position of hexagons are designed to cover the `core' region (top panels of Figure\,\ref{fig:figure7}). The peak position of the [O\,{\sc iii}]\,5007\AA\, line is matched with the central position of Region 6.

From each region, we extracted the spectra and derived the gas-phase metallicity, SFR surface density, electron density, and ionization parameter following the same approach outlined in Section\,\ref{subsec:section3.2}. In Figure\,\ref{fig:figure7}, we compare these physical properties to investigate the sub-kpc variation within this system, particularly as a function of the SFR surface density. The gas-phase metallicity is constrained in all regions except Regions 3 and 6. All of the regions have lower metallicity than the `tail' region (panel a), consistent as our previous finding in Section\,\ref{subsec:section3.2.2}. Similarly, most regions have ionization parameters comparable to that of the `tail' region , with the exception of Regions 3 and 6 (panel c and d). These results indicate that the ISM conditions in Regions 3 and 6 have distinct properties, characterized by higher ionization parameter compared to other regions . 

Due to the low S/N of the \oii doublet, unfortunately, all of the electron densities are unconstrained, providing either upper or lower limits. However, we find a trend that the regions with higher SFR surface density ($\log(\Sigma_{\rm SFR}/M_{\odot}\,{\rm yr}^{-1}\,{\rm kpc}^{-2})\gtrsim0.8$) tend to have higher electron density and vise versa (panel b). This is qualitatively consistent with the correlation found by \cite{2025ApJ...993L..51L}, but likely with a steeper slope. Although we need higher S/N to constrain the slope of this $\Sigma_{\rm SFR}-n_{e}$ correlation, our finding suggests the same physical phenomenon is present in $z>9$ galaxies; that is, active star formation occurs in high-density gas environments, resulting in strong stellar feedback and high ambient pressure, which in turn lead to high electron densities. 

We do not find a clear correlation between SFR surface density and ionization parameter (panel c). The `core' region as a whole shows a good agreement with the correlation made from star-forming galaxies at $1.9<z<3.7$ \citep{2023ApJ...951...56R}. However, the spatially resolved ionization parameter exhibits a large scatter, with two regions (Regions 3 and 6) lying more than 0.2\,dex above the correlation and the remaining regions lying below it. This highlights the variation in ionized gas conditions in Gz9p3 is pronounced (up to 0.5\,dex) when resolved down to sub-kpc scales, which was not seen at kpc scales (Section\,\ref{subsec:section3.2.4}). 

Panel (d) of Figure\,\ref{fig:figure7} compares the metallicity and ionization parameter, where we found a positive correlation. This is somewhat inconsistent with the anti-correlation commonly reported in the literature, such that low-metallicity galaxies have higher ionization parameter \citep{2019ARA&A..57..511K}. However, correlations between these two parameters are seen in nearby galaxies, particularly when galaxies are spatially-resolved (e.g., \citealp{2018MNRAS.479.5235P}; \citealp{2022A&A...659A.112J}), although the correlation is dependent on adopted photoionization models and its physical origin remains uncertain \citep{2022A&A...659A.112J}. By exploiting the power of NIRSpec IFU, more statistical investigations on the spatially-resolved correlation between ionization parameter and metallicity are needed. 

\subsubsection{Discovery of ionized outflows} 
\label{subsec:section3.3.1}

By inspecting the [O\,{\sc iii}]\,5007\AA\,\,line spectra, we found that some regions, including the `tail' spectrum, exhibit either symmetric (e.g., Regions 3, 5 and 6) or asymmetric (e.g., Regions 1 and 2) broad wing components. As no evidence of AGN is found in this system (\citealp{2024NatAs...8..657B}; \citealp{2026arXiv260422460B}), we attribute the origin of these broad \oiii components as star formation-driven outflows.

We perform two Gaussian component fittings to characterize these outflow components and present the fitting results in Figure\,\ref{fig:figure8}. We estimate the outflow velocity $v_{\rm out}$ as 
\begin{equation}
v_{\rm out} = \Delta v + 2\sigma_{\rm broad} \,\,\, (\Delta v = |v_{\rm narrow}-v_{\rm broad}|),
\end{equation}
which represent the summation of the velocity offset between narrow and broad components ($\Delta v$) and the width of broad component ($\sigma_{\rm broad}$) . Region 6 and ambient regions (Regions 3 and 5) show a broad wing with small velocity offset ($\Delta v\lesssim 50$\,km\,s$^{-1}$) and therefore moderate outflow velocity ($v_{\rm out}\sim 100-150$\,km\,s$^{-1}$). These outflow velocities are consistent with the values reported by \cite{2026arXiv260421516C} and \cite{2026arXiv260421218Z} from the UV absorption lines. This suggests that the structure of outflows blowing out from the central core are isotropic (at least along the line of sight). On the other hand, regions toward the northeast exhibits prominent redshifted component ($\Delta v \sim 200-250$\,km\,s$^{-1}$), resulting in high outflow velocity ($v_{\rm out}\sim 350-400$\,km\,s$^{-1}$). The outflow velocities in the northeast regions are comparable to other high-redshift galaxies with ionized outflows (e.g., \citealp{2024A&A...685A..99C}; \citealp{2025MNRAS.544.4722R}; \citealp{2025ApJ...984..182X};  \citealp{2026A&A...711A..66R}). 

Whether these outflows escape the gravitational potential of this system is crucial in the context of chemical enrichment of circumgalactic medium (CGM) and baryon cycle. Nonetheless, it is challenging to estimate the escape velocity ($v_{\rm esc}$) of Gz9p3, because of its complex morphology and the lack of its dynamical mass. Therefore, as a first-order approximation, we estimate the escape velocity using the halo mass of Gz9p3 from the stellar-to-halo mass relation derived by \cite{2025A&A...695A..20S} and $v_{\rm esc}\sim3v_{\rm cir}=3(GM_{\rm h}/r_{\rm h})^{1/2}$ following \cite{2025ApJ...984..182X}, where $r_{\rm h}$ is the virial radius of the dark matter halo \citep{1996MNRAS.282..347M}. At $8.5<z<10$, typical galaxies with $M_{\star}\sim 10^{9}M_{\odot}$ are derived to reside in dark matter halos with $M_{\rm h}\sim8\times10^{10}M_{\odot}$ \citep{2025A&A...695A..20S}, leading to the escape velocity estimate of $v_{\rm esc}\sim490$\,km\,s$^{-1}$. 

Given that the maximum outflow velocity we obtained from Gz9p3 is $v_{\rm out} = 399$\,km\,s$^{-1}$, we found that the $v_{\rm out}/v_{\rm esc}$ ratio in Gz9p3 is below unity ($\lesssim0.8$). This indicates that the outflowing gas is likely gravitationally bound to Gz9p3 and therefore does not efficiently enrich its CGM on scales beyond a few kpc. Instead, the outflowing gas will likely return to the ISM of Gz9p3 due to the gravitational force. This marks Gz9p3 as one of the first galaxies experiencing a galactic fountain.

\section{Discussion and Summary}
\label{sec:section4}

\begin{figure}
	\center
    \includegraphics[width=0.96\linewidth]{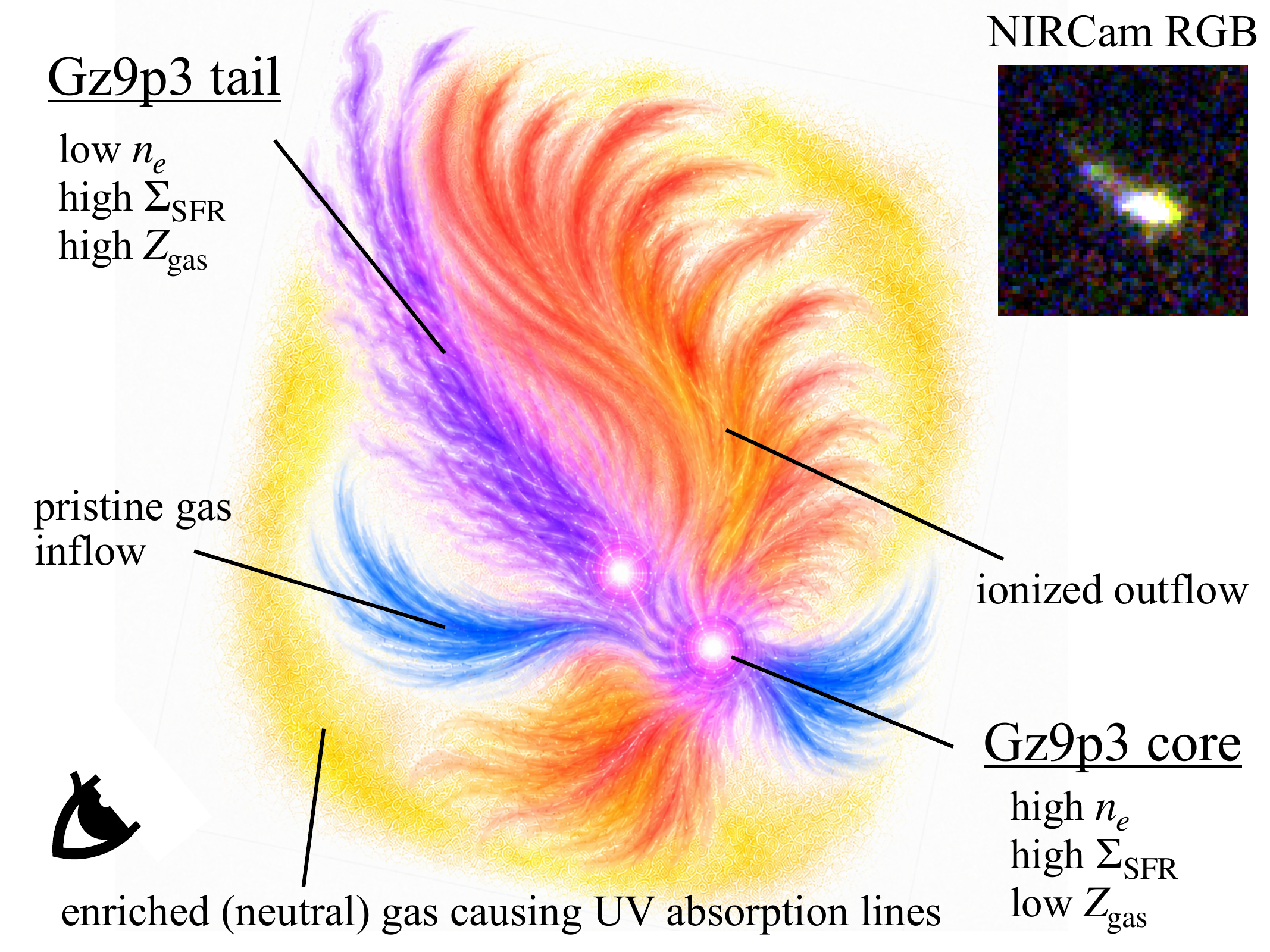}
    \caption{Schematic illustration of Gz9p3. Purple regions depict the distributions of stars and star formation. Red and blue indicate the [O\,{\sc iii}] outflows and metal-poor inflow, respectively. On the top right corner, we display the NIRCam RGB image of Gz9p3. The illustration was first hand-drawn by the authors and then refined using ChatGPT.}
    \label{fig:figure9}
\end{figure}

We have investigated the physical properties of Gz9p3, one of the most extended systems confirmed at $z>9$. Our major findings are as follows: 1) the `core' region exhibits a deficit in oxygen abundance ($\sim0.25-0.5$\,dex below the MZR) than the `tail' region,  2) star formation is active in the `core' (${\rm SFR}=23.8\,M_{\odot}\,{\rm yr^{-1}}$) but its sSFR is lower than the `tail', 3) the system has a wide range of electron densities (across more than 1.5\,dex of magnitude) and 4) ionized outflows ($v_{\rm out}/v_{\rm esc} <1$) are present from the `core' to the `tail' regions. Taken together, we present the schematic picture of Gz9p3 in Figure\,\ref{fig:figure9}, which potentially explains all of the observational facts about this system. 

The deficit of gas-phase metallicity in the `core' region leads to an anti-correlated MZR (Figure\,\ref{fig:figure5}) at a kpc scale. From the spatially-resolved SED fitting analysis, \cite{2024NatAs...8..657B} report that the stellar ages of Gz9p3 are homogeneous ($\sim25$\,Myr) with no significant variation, which would result in a flat metallicity distribution. Therefore, one scenario which may explain this deficit is an infall of metal-poor/pristine gas from the CGM to the central core of Gz9p3. As the stellar mass is dominated by the `core' region, it can be explained that the pristine gas from the CGM preferentially falls into the `core' region. Similar discussions has been made by \cite{2024MNRAS.533.2488M} for MACS1149-JD1, although they explain the deficit by gas replenishment through gas inflow within the system. Given that the `core' region also expels gas by ionized outflows, we suspect that the metallicity of the `core' region is diluted by inflows anisotropically, as highlighted in Figure\,\ref{fig:figure9}. 

The intense star formation is present both in the `core' ($\Sigma_{\rm SFR}=5.9\,M_{\odot}{\rm yr}^{-1}\,{\rm kpc}^{-2}$) and `tail' ($\Sigma_{\rm SFR}=4.5\,M_{\odot}{\rm yr}^{-1}\,{\rm kpc}^{-2}$) regions, which is likely caused by galaxy interactions in the past. Combined with young stellar ages, Gz9p3 is consistent with the properties of merger-driven star-forming galaxies reproduced in the cosmological zoom-in simulations \citep{2024ApJ...975..238N}. The most extreme ISM conditions are found at the peak position of the [O\,{\sc iii}]\,5007\AA\, line emission (Region 6 in Figure\,\ref{fig:figure7}) with $\log (\Sigma_{\rm SFR}/M_{\odot}{\rm yr}^{-1}\,{\rm kpc}^{-2})=0.42\pm0.03$, $\log U=-2.11_{-0.10}^{+0.07}$, and $n_{e}>2100$\,cm$^{-3}$. However, we note that these are not as extreme as UV-bright, compact galaxies at higher redshift ($z\gtrsim10$). For example, GHZ2 at $z=12.34$ exhibits much higher SFR surface density $\log (\Sigma_{\rm SFR}/M_{\odot}{\rm yr}^{-1}\,{\rm kpc}^{-2})=2.11^{+0.35}_{-0.25}$, ionization parameter $\log U=-1.75\pm0.16$ (\citealp{2024ApJ...972..143C}; \citealp{2024ApJ...975..245C}), and electron density $n_{e}\lesssim4200$\,cm$^{-3}$ (constrained from two \oiii lines at FIR wavelengths; \citealp{2024ApJ...977L...9Z}). This indicates that galaxy interactions at this epoch cannot reproduce such extreme conditions as those observed in GHZ2 and other high-redshift galaxies (see e.g., \citealp{2026arXiv260708749L} for recent compilation). 

What is puzzling is the origin of diverse electron density ($\gtrsim1.5$\,dex variation) in Gz9p3, which significantly deviates from the empirical correlation (Figures\,\ref{fig:figure6} and \ref{fig:figure7}). Such large variation of electron density at similar redshifts is also reported in the literature (\citealp{2024MNRAS.533.2488M}; \citealp{2025ApJ...985...83M}; \citealp{2026arXiv260514922T}), and have largely been discussed in the context of redshift evolution of sizes and metallcity (e.g., \citealp{2023ApJ...956..139I}; \citealp{2024ApJ...973...47A}). However, such diverse scatter of electron density within a single system cannot be explained by redshift evolution. Some other studies attributed high electron densities at high redshift to the strong ISM pressure from stellar winds or ambient medium, and the elevated star formation activity governed by the density of molecular clouds (e.g., \citealp{2015MNRAS.451.1284S}; \citealp{2017MNRAS.465.3220K}; \citealp{2021ApJ...909...78D}).

In particular, \cite{2021ApJ...909...78D} suggested that the electron densities of star-forming galaxies at $0<z<2.6$ linearly correlate with molecular gas densities ($n_{\rm H_{2}}$) and the trend can plausibly be explained by the gas densities of the parent molecular clouds averaged over more than a few hundred pc scales. In other words, the electron density is roughly proportional to the molecular surface density ($\Sigma_{\rm mol \ gas}$), thus the slope of the $n_{e}$-$\Sigma_{\rm SFR}$ correlation can be regarded as an inverse of the slope of Kennicutt-Schmidt (KS) law \citep{1998ApJ...498..541K}. The $n_{e}-\Sigma_{\rm SFR}$ correlations reported by \cite{2023ApJ...951...56R} and \cite{2025ApJ...993L..51L} have a slope of $\sim0.5$, which can be translated into a KS slope of $\sim2$, although other factors such as ambient gas pressure and diffuse ionized gas can influence the slope. If Gz9p3 truly has steeper slope in the $n_{e}$-$\Sigma_{\rm SFR}$ correlation (Figure\,\ref{fig:figure7}), this suggests that the KS slope is shallower than other galaxies, i.e., star formation is inefficient at dense regions. Given that the `core' region is characterized by a post-burst star formation history \citep{2026arXiv260421516C}, it is possible that there exists a cold neutral gas which is not efficiently used as a supply of star formation activity, due to strong stellar feedback and star formation-driven outflows.


The largely missing piece in this discussion is the information of cold neutral gas associated to the ISM of Gz9p3, which is indirectly detected from UV absorption lines (e.g., \citealp{2026arXiv260421516C}; \citealp{2026arXiv260421218Z}). 
Future observations of gas tracers such as [C\,{\sc ii}]\,158$\mu$m line which can be observed by ALMA, is awaited for assessing the star formation law and star formation efficiency in a spatially resolved manner.

In this work, we presented the JWST NIRSpec/G395H IFU observation of Gz9p3, an interacting galaxy consisting of the `core' and `tail' sub-regions. From the rest-frame optical lines ([O\,{\sc ii}]3727,\,3729, H$\beta$, [O\,{\sc iii}]\,4959,\,5007\AA), we found large scatters in the ISM conditions, particularly in electron density and gas-phase metallicity. As this galaxy has likely been experiencing a galaxy interaction, we argue that this galaxy is in a dynamic, dispersion-dominated phase where both inflows and outflows are playing a crucial role. The metal-enriched CGM has been suggested from the UV absorption lines (\citealp{2026arXiv260421516C}; \citealp{2026arXiv260421218Z}), but our understanding of this system may still be incomplete and biased toward the properties of the `core' region. Future JWST IFU studies combining shorter-wavelength gratings will be crucial for confirming the geometry of pristine gas, outflows, and the enriched CGM. 

\section*{Acknowledgements}


TM acknowledges support for this work by JSPS KAKENHI Grant Number K25K24918. MS acknowledges support for this work under NASA grant 80NSSC22K1294. AC acknowledges support from the Next Generation EU-funded project IR0000034, ``STILES -- Strengthening the Italian Leadership in ELT and SKA''.

This work is based in part on observations made with the NASA/ESA/CSA James Webb Space Telescope. The data were obtained from the Mikulski Archive for Space Telescopes (MAST) at the Space Telescope Science Institute, which is operated by the Association of Universities for Research in Astronomy, Inc., under NASA contract NAS 5-03127 for JWST. These observations are associated with program \#\,4553. Furthermore, this paper makes use of the following ALMA data: ADS/JAO.ALMA \#\,2024.1.00440.S. ALMA is a partnership of ESO (representing its member states), NSF (USA) and NINS (Japan), together with NRC (Canada), MOST and ASIAA (Taiwan), and KASI (Republic of Korea), in cooperation with the Republic of Chile. The Joint ALMA Observatory is operated by ESO, AUI/NRAO and NAOJ. We thank the ALMA staff and in particular the EA-ARC staff for their support. Data analyses were in part carried out on the Multi-wavelength Data Analysis System operated by the Astronomy Data Center (ADC), National Astronomical Observatory of Japan. The authors would also like to acknowledge the use of AI language model ChatGPT (GPT-5.6 Luna; \url{https://chatgpt.com}) for assistance in refining the writing and language of this manuscript, and in generating a schematic illustration.

\section*{Data Availability}

The data are publicly available through the Mikulski Archive for Space Telescopes (MAST) portal managed by Space Telescope Science Institute.



\bibliographystyle{mnras}
\bibliography{arxiv_v1} 




\newpage
\appendix

\section{Spectral Fitting and Line Flux Measurements}
\label{sec:sectionA}

\begin{figure*}
	\includegraphics[width=0.85\linewidth]{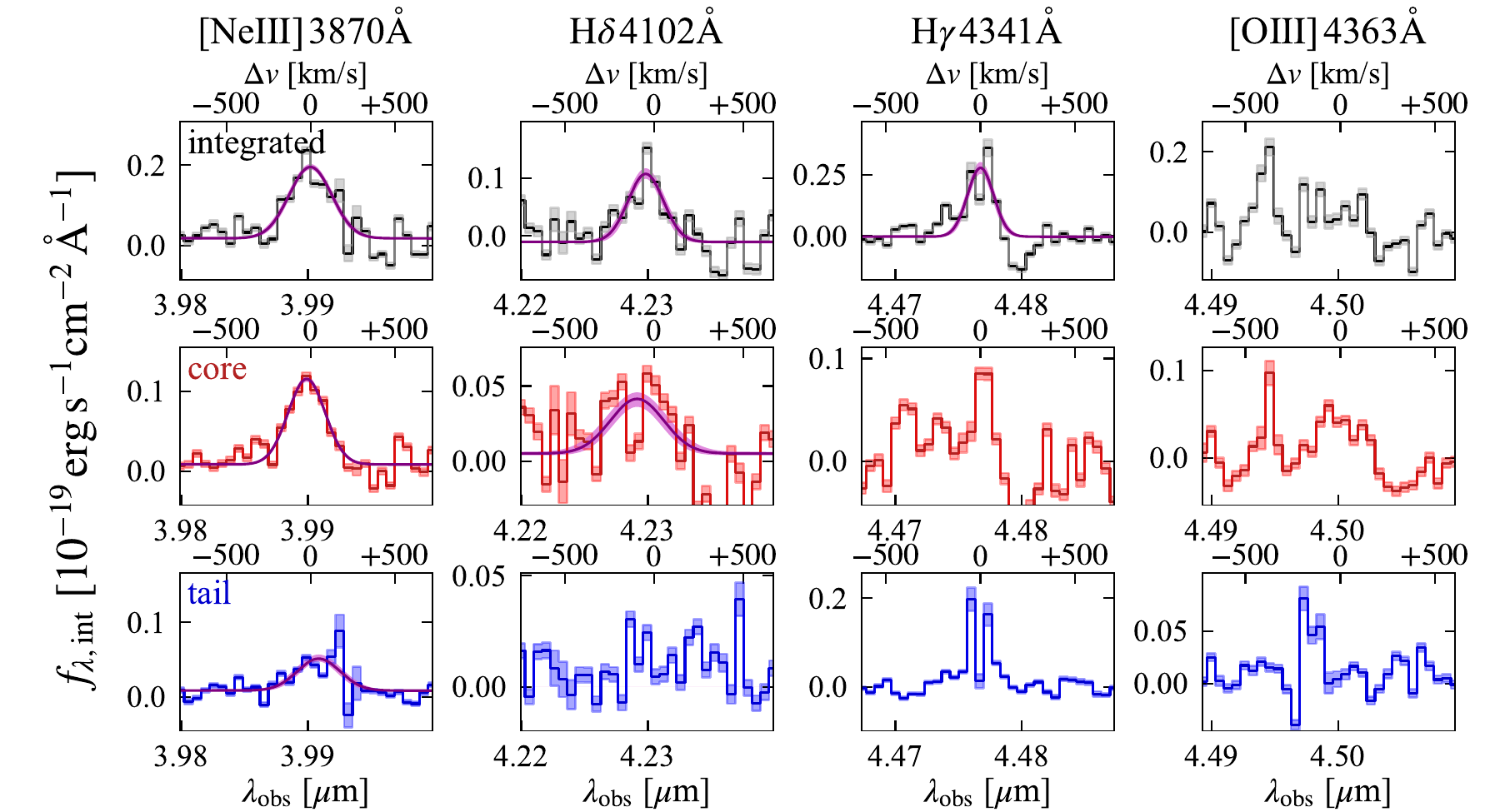}
    \caption{Same as Figure\,\ref{fig:figure3}, but for other emission lines useful for the line diagnostics. Best-fit Gaussian fittings are shown in the purple line only for those detected at $>3\sigma$ significance. }
    \label{fig:figureA1}
\end{figure*}

To measure the line fluxes, we performed spectral fitting using a model consisting of a single Gaussian profile and a constant continuum using the {\tt scipy/curve\_fit} package except for the \oii doublet. We added random noise to the extracted spectra drawn from the ERR extension map, using it as the standard deviation, and repeated the fitting procedure 1000 times. We then adopted the median value as the measured flux, and the 16th and 84th percentiles as the uncertainties. 

For lines that are not detected above $3\sigma$ significance, we estimated the upper limit of line fluxes as follows. We first calculate 1) the integrated line flux directly from the extracted spectra, 2) the summation of the errors in quadrature ($\sqrt{\sum_{i}\sigma_{i,{\rm err}}^{2}}$), and  3) if available, the flux uncertainties derived from the Gaussian model fitting. Here, we use the spectral channels within $\lesssim125$\,km\,s$^{-1}$ of the redshifted line center, assuming a line width of 250\,km\,s$^{-1}$. We conservatively adopt the maximum value among these three uncertainties and report $3\sigma$ upper limits in Table\,\ref{tab:tablea1}. 

\begin{table}
	\centering
	\caption{Line fluxes and line widths (FWHM) for both the integrated region and the subregions of Gz9p3.}
	\label{tab:tablea1}
    \renewcommand{\arraystretch}{1.5}
    \setlength{\tabcolsep}{1.5pt}
	\begin{tabular}{lcccccc} 
		\hline
		line & \multicolumn{2}{c}{integrated} & \multicolumn{2}{c}{core} & \multicolumn{2}{c}{tail} \\
        \cmidrule(lr){2-3} \cmidrule(lr){4-5} \cmidrule(lr){6-7}
        & flux & FWHM & flux & FWHM & flux & FWHM \\
		\hline
        [O\,{\textsc{ii}}]\,$\lambda$3726 & $6.13_{-0.24}^{+0.23}$ & $230_{-10}^{+12}$ & $3.72_{-0.27}^{+0.38}$ & $270_{-15}^{+18}$ & $1.48_{-0.08}^{+0.07}$ & $193_{-8}^{+36}$\\
        
        [O\,{\textsc{ii}}]\,$\lambda$3729 & $8.38_{-0.29}^{+0.36}$ & -- & $2.14_{-0.30}^{+0.23}$ & -- & $2.17_{-0.07}^{+0.19}$ & -- \\
        
        [Ne\,{\textsc{iii}}]\,$\lambda$3870 & $7.51_{-0.35}^{+0.34}$ & $299_{-16}^{+17}$ & $3.91_{-0.11}^{+0.12}$ & $259_{-8}^{+8}$ & $1.84_{-0.22}^{+0.15}$ & $298_{-31}^{+26}$ \\
        
        H$\delta$\,$\lambda$4102 & $4.11_{-0.30}^{+0.26}$ & $233_{-24}^{+20}$ & $1.94_{-0.18}^{+0.22}$ & $351_{-26}^{+44}$ & $<1.36$ & -- \\
        
        H$\gamma$\,$\lambda$4341 & $7.24_{-0.55}^{+0.52}$ & $162_{-14}^{+13}$ & $< 8.89$ & -- & $< 9.68$ & -- \\
        
        [O\,{\textsc{iii}}]\,$\lambda$4363 & $< 9.56$ & -- & $< 2.38$ & -- & $< 1.49$ & -- \\
        
        H$\beta$\,$\lambda$4861 & $15.21_{-0.42}^{+0.41}$ & $277_{-10}^{+10}$ & $7.38_{-0.28}^{+0.29}$ & $316_{-15}^{+15}$ & $2.81_{-0.17}^{+0.22}$ & $222_{-20}^{+27}$\\
        
        [O\,{\textsc{iii}}]\,$\lambda$4959 & $31.07_{-0.44}^{+0.46}$ & $225_{-5}^{+5}$ & $14.28_{-0.25}^{+0.24}$ & $222_{-5}^{+5}$ & $6.80_{-0.17}^{+0.16}$ & $205_{-6}^{+7}$ \\
        
        [O\,{\textsc{iii}}]\,$\lambda$5007 & $75.40_{-0.62}^{+0.43}$ & $182_{-1}^{+2}$ & $37.34_{-0.27}^{+0.27}$ & $187_{-2}^{+1}$ & $14.90_{-0.16}^{+0.17}$ & $174_{-2}^{+1}$ \\
        \hline
	\end{tabular}
    \begin{tablenotes}
    \item All of the fluxes are corrected for lensing magnification, but not for dust attenuation. Line fluxes are in units of $10^{-19}\,{\rm erg\,s^{-1}\,cm^{-2}}$. FWHMs are in units of km\,s$^{-1}$ and not corrected for the instumental broadening.
    \end{tablenotes}
\end{table}

\subsection{Double Gaussian fitting of the \oii doublet}
\label{subsec:sectionA1}

\begin{figure*}
	\includegraphics[width=0.85\linewidth]{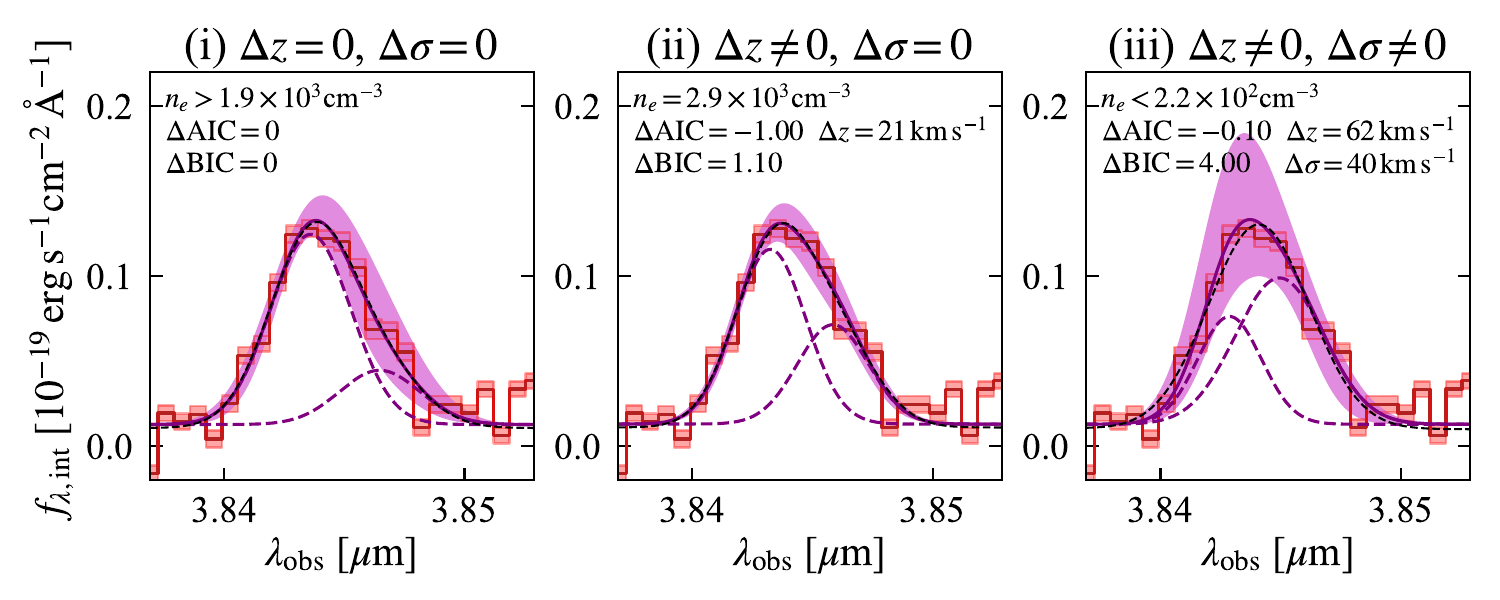}
    \caption{Comparison of the double Gaussian fittings for the \oii doublet in the `core' region. Each fitting has  different set of free parameters: (i) the doublet is constrained to have the same redshift ($\Delta z=0$) and line width ($\Delta \sigma=0$), (ii) the doublet is allowed to have different redshifts ($\Delta z\neq0$) but are constrained to have the same line width ($\Delta \sigma=0$), and (iii) the doublet is allowed to have different redshifts ($\Delta z\neq0$) and different line widths ($\Delta \sigma\neq0$). The fittings were performed in the same manner as other lines (Appendix\,\ref{sec:sectionA}). The black dashed line represents the best-fit model obtained from a direct fit to the spectral data (without adding random noise) and is used to calculate the AIC and BIC values. The derived electron density and the differences in AIC and BIC with respect to fit (i) are shown in the upper-left corner. The redshift and line width differences between the two \oii lines for fits (ii) and (iii) are expressed in units of velocity.}
    \label{fig:figureA2}
\end{figure*}

For the \oii doublet, we perform a double Gaussian and constant continuum fitting. We set the doublet to  have the same FWHM and redshift. However, for the `core' region, the fitted doublet provides significantly low [O\,{\sc ii}]\,3729\AA/[O\,{\sc ii}]\,3726\AA\, line ratio of $0.28^{+0.15}_{ -0.11}$, converted into the electron density of $n_{e}>1.3\times10^{4}$\,cm$^{-3}$. This electron density well exceeds the highest values reported at high redshift (\citealp{2023ApJ...956..139I}; \citealp{2024ApJ...973...47A}) and lies in a regime where the \oii line ratio is largely insensitive to electron density \citep{2016ApJ...816...23S}. We consider this result to be unreliable, likely owing to the limited spectral resolution and contamination from various line-of-sight components. Therefore, we allow the doublet to have different redshifts ($\Delta z = z_{\rm [OII]3729} - z_{\rm [OII]3726} \neq 0$) or different line widths ($\Delta \sigma = \sigma_{\rm [OII]3729} - \sigma_{\rm [OII]3726} \neq 0$). Figure\,\ref{fig:figureA2} compares three fitting results obtained with different sets of free parameters: (i) the doublet is constrained to have the same redshift and line width, (ii) the doublet is allowed to have different redshifts but are constrained to have the same line width, and (iii) the doublet is allowed to have different redshifts and different line widths. To statistically assess which model is preferred, we calculate the Akaike Information Criterion (AIC) and Bayesian Information Criterion (BIC), both of which account for the number of free parameters in the fit. In each panel, the differences of AIC and BIC relative to the fit (i) are shown; smaller values indicate a stronger preference for the model.

We find that both fits (ii) and (iii) are preferred over fit (i) according to the AIC, although the differences are not statistically significant ($\Delta{\rm AIC}<-2$). The BIC does not favor fit (iii). Considering these results, we decided to report the fit (ii) as our measurements, providing the electron density $n_{e}=2870^{+1150}_{-1370}$\,cm$^{-3}$.
The redshift offset between the two \oii lines is 21\,km\,s$^{-1}$ in velocity, which is below the velocity resolution of NIRSpec G395H/F290LP ($\approx110$\,km\,s$^{-1}$).

\bsp	
\label{lastpage}
\end{document}